\documentclass[a4paper,fleqn,usenatbib]{mnras}
\pdfoutput=1

\usepackage[T1]{fontenc}
\usepackage{ae,aecompl}

\usepackage{graphicx}	
\usepackage{amsmath}	
\usepackage{amssymb}	

\usepackage[normalem]{ulem}
\usepackage{xcolor,cancel}
\usepackage{comment}

\title[New discontinuous, 1D solar wind solutions]{A new class of
discontinuous solar wind solutions}

\author[B. M. Shergelashvili et al.]{
Bidzina M. Shergelashvili,$^{1,2,3}$\thanks{E-mail: bidzina.shergelashvili@oeaw.ac.at or bidzina.shergelashvili@iliauni.edu.ge}
Velentin N. Melnik,$^{4}$
Grigol Dididze,$^{2,3,5}$ 
\newauthor{
Horst Fichtner,$^{6}$
G\"{u}nter Brenn,$^{7}$
Stefaan Poedts,$^{5,8}$
Holger Foysi,$^{9}$}
\newauthor{
Maxim L. Khodachenko,$^{1,10,11}$
and Teimuraz V. Zaqarashvili$^{12,13,3}$}
\\
$^{1}$Space Research Institute, Austrian Academy of Sciences, Schmiedlstrasse
      6, 8042 Graz, Austria \\
$^{2}$Centre for Computational Helio Studies at Faculty of Natural Sciences and Medicine, Ilia State University, \\
      Cholokashvili Ave. 3/5, 0162 Tbilisi, Georgia \\
$^{3}$Evgeni Kharadze Georgian National Astrophysical Observatory, Abastumani, 0301 Adigeni Municipality, Georgia \\
$^{4}$Institute of Radio Astronomy, National Academy of Sciences of
      Ukraine, Kharkov, Ukraine \\
$^{5}$Centre for mathematical Plasma Astrophysics, KU Leuven,
      Celestijnenlaan 200 B, B-3001, Leuven, Belgium \\
$^{6}$Institut f\"ur Theoretische Physik IV, Plasma-Astroteilchenphysik,
      Universit\"atsstra{\ss}e 150, 44801 Bochum, Germany \\
$^{7}$Institute of Fluid Mechanics and Heat Transfer, Graz University of
      Technology, 8010 Graz, Austria \\
$^{8}$Institute of Physics, University of Maria Curie-Sk{\l}odowska, PL-20-031 Lublin, Poland \\
$^{9}$Institut f\"ur Fluid- und Thermodynamik, Universit\"at Siegen,
      Paul-Bonatz-Str. 9-11, 57068 Siegen, Germany \\
$^{10}$Skobeltsyn Institute of Nuclear Physics, Moscow State University,
      Moscow, Russia \\
$^{11}$Institute of Laser Physics, SB RAS, Novosibirsk 630090, Russia\\
$^{12}$Institute of Physics, IGAM, University of Graz,
      Universit\"atsplatz 5, 8010 Graz, Austria \\
$^{13}$Faculty of Natural Sciences and Medicine, Ilia State University,
      Cholokashvili Ave. 3/5, 0162 Tbilisi, Georgia \\
}

\date{Accepted XXX. Received YYY; in original form ZZZ}

\pubyear{2020}

\begin{document}
\label{firstpage}
\pagerange{\pageref{firstpage}--\pageref{lastpage}}
\maketitle

\begin{abstract}
A new class of one-dimensional solar wind models is developed within the general polytropic, single-fluid hydrodynamic 
framework. The particular case of quasi-adiabatic radial expansion with a localized heating source is considered. We consider analytical solutions with continuous
Mach number over the entire radial domain while allowing for jumps in the flow velocity, density, and temperature, provided that there exists an external source of energy in the vicinity of the critical point which supports such jumps in physical quantities.
This is substantially distinct from both the standard Parker solar wind model and the original nozzle solutions, where
such discontinuous solutions are not permissible. We obtain novel sample analytic solutions of the governing equations 
corresponding to both slow and fast wind. 
\end{abstract}




\section{Introduction}

The modeling of the quasi-stationary patterns of the solar wind flows
originated from the pioneering work of Eugene Parker establishing the 
concept of the solar wind \citep{parker1958} and its applications 
to its interaction with the tererestrial magnetosphere \citep{Parker-1958b}, 
to magnetic field instabilities \citep{Parker-1958c}, 
to cosmic ray modulation \citep{Parker-1958d, Parker-1965}, 
and to shocks \citep{Parker-1959a, Parker-1959b}. The 
fundamental importance of such studies for our understanding of the physics of
the Sun, its planetary system and the entire heliosphere has been proven during
the past six decades by numerous investigations. There is a particular need
to investigate the basic mechanisms that drive the dynamics and thermodynamics
of the solar atmosphere and the wind outflow into the interplanetary space,
thus shaping the whole heliosphere. The vast progress made in both analytical
and numerical modelling \citep[see, e.g., the reviews][]{Tsinganos-2007, 
Cranmer-Winebarger-2019} has been accompanied by an extensive collection of
observational data from a wide  variety of space and ground based missions.
Currently, state-of-the-art solar and space condition monitoring infrastructures
exist (via the spacecraft ACE, SDO, Parker Solar Probe etc.) as well as
data-driven prediction and research models, like AWSoM \citep{vanderholst14},
ENLIL \citep{enlil2003}, or EUHFORIA \citep{pomoell18}.

As such, modern space weather research infrastructures consist of combinations
of many analytical, numerical and observational methodologies that complement
each other. The mutual development of these three types of approaches is one
of the main challenges, motivated by the need for a deeper insight in
the physical mechanisms behind the solar/space weather, planetary and/or
heliospheric physics. The content of the present paper falls in this framework
and represents an attempt to contribute to the described challenges. We refer
the interested reader to the well-structured reviews on the nature of the solar
wind sources in the corona/solar atmosphere \citep{feldman2005} and solar wind
interplanetary patterns \citep{enhim2011}, and references therein. 

One of the important aspects in space weather is the physical
state of the solar corona and of the inner solar atmosphere where the physical 
driving processes of different flow patterns are located. The
identification and classification of these sources is possible through remote
observations of the processes in the vicinity of the Sun (e.g., 
electromagnetic waves in the radio range) and in-situ data at the Lagrangian
(L1) point, in combination with analytical and numerical models as conceptual
ground for interpretation and analysis of these data.

In the present paper, we consider a one-dimensional stationary model for the radial profiles
of the physical quantities in the non-isothermal, quasi-polytropic solar wind
patterns. We derive a class of solutions that are obtained from the hydrodynamic
1D model of the Solar Wind (SW) where polytropic relations between the pressure,
density and temperature hold \citep{jacobs2011} outside of discontinuities.
The single-fluid description is chosen to reveal 
the core of the physical nature of the new solutions and their behaviour at the critical 
point. We address the crucial issue of 'integration constants' that represent (dynamic)
invariants of the profile combinations. The formalism follows the same philosophy
as the standard Parker's wind mode, i.e.\ both the isothermal one \citep{parker1958} or 
the later model including the effect of the thermal conduction \citep{parker1965b}. 

In spite of the fact that the novel solutions are derived within in a general
polytropic framework, we discuss throughout this paper the particular case of adiabatic
systems where there is no exchange of heat between the flowing fluid elements
and the environment and any changes of the internal
energy of the plasma occur only at the expense of the work
done by a plasma parcel on its surroundings or vise versa. 

We emphasise that the model presented here is stationary and purely radial.
The latter condition implies that, similarly to the standard Parker's hydrodynamic
SW model, it is assumed that the flow profile is formed along an irrotational,
spherically symmetric radial magnetic field with the consequence that the influence
of the Lorentz force on the plasma flow is excluded. Such 1D models mimic only
roughly the complete patterns of the SW flow from the Sun to $1\;$AU and beyond.
However, their physical precision is acceptable within the altitude ranges where
the magnetic and flow fields 
rotate rigidly,
i.e.\ the region with $r<0.1\;$AU, which is the subject of main interest here.
Nevertheless, one must be aware that such hydrodynamic descriptions are used as
a first approximation 
of the shape and dynamics of the inner heliosphere.

Following a description of the model in section~2, we give an interpretation of the 
transsonic solutions in section~3 with an emphasis on a new class of solutions, give 
examples for the latter, and discuss the required physical conditions.
Finally, in section~4 we summarize the results and draw conclusions.

\section{Description of the model} \label{sec:model}
The model is based on the following equations:
\begin{equation}
\frac{\partial\rho}{\partial
t}+\mathbf{\nabla}\cdot\rho\mathbf{v}=0,\label{eq:contin}
\end{equation}
\begin{equation}
\rho\left[\frac{\partial\mathbf{v}}{\partial
t}+\left(\mathbf{v}\mathbf{\cdot\nabla}\right)\mathbf{v}\right]=-\nabla
p-\rho\frac{GM_{\sun}}{r^{2}}\cdot\mathbf{n_{r}},
\label{eq:momentum}
\end{equation}
where, $\rho$, $p$, $\mathbf{v}$, and $\mathbf{B}$ denote the plasma mass density,
(thermal) pressure, velocity, respectively. $G$ is the
gravitational constant, $M_{\sun}$ the mass of the Sun, and
$\mathbf{n_{r}}$ a radial unit vector. Furthermore, we
consider the energy equation of a polytropic gas flow,
which is valid for the case of isentropic (adiabatic) flow patterns and, along
with the ideal gas law, closes the system of the governing equations:
\begin{eqnarray}
\left[\frac{\partial}{\partial
t}+\left(\mathbf{v}\mathbf{\cdot\nabla}\right)\right]\left(\frac{T}{\rho^{
\alpha-1}}\right)= 0.
\label{eq:energy}
\end{eqnarray}
with $\alpha$ denoting the polytropic index (we will consider here the case
where $\alpha=\gamma$, where $\gamma$ is the adiabatic index).
In general, on the r.h.s.\ of this equation source terms appear, encompassing
the heat (entropy) production or transport that can maintain the irreversible
polytropic motion of plasma elements. Here, however, we limit the study to 
adiabatic processes, with the aim to reveal fundamental properties of the 
new class of stationary solutions for the flow patterns. 

For the case of purely radial but non-isothermal and stationary outflow (possibly
along a global radial magnetic field, so that the Lorentz force vanishes) one has: 
\begin{equation}
r^{2}\rho v=Q_{\rho}=\mathrm{const},\label{eq:radcont}
\end{equation}
\begin{equation}
v\frac{dv}{dr}=-\frac{1}{\rho}\frac{dp}{dr}-\frac{GM_{\sun}}{r^{2}},\label{eq:radmoment}
\end{equation}
\begin{equation}
\frac{d}{dr}\frac{T}{\rho^{\alpha-1}}=0,\label{eq:radenergy}
\end{equation}
where, similarly to the standard isothermal Parker solar wind outflow,
the radial mass flow rate $Q_{\rho}$ is an integration constant determined by
the boundary condition at a given location along the radius (in  standard cases
it is used to satisfy values of the density and flow velocity at 1~AU or at the 
so-called source surface, where the wind is assumed to be purely radial). 
$v$ denotes the radial component of the velocity, $\rho=\mu m_{p}n$ is the plasma
mass density with the mean molecular weight $\mu m_{p}$
(where $\mu$ is the mean molecular mass in units of the proton mass
$m_{p}$ and $n$ is the plasma number density).
Conveniently, the ideal gas law holds for the considered flow patterns
$p=nk_{B}T$, where $k_{B}$ is the Boltzmann
constant. With the ideal gas law one obtains from Eqs.~(\ref{eq:radmoment}) and
(\ref{eq:radenergy}):
\begin{equation}
v\frac{dv}{dr}=-C_{s}^{2}\frac{1}{\rho}\frac{d\rho}{dr}-\frac{GM_{\sun}}{r^{2}},
\label{eq:radmoment3}
\end{equation}
where,
\begin{equation}
C_{s}^{2}=\left(\frac{\partial p}{\partial\rho}\right)_S=\frac{\alpha
k_{B}T}{\mu
m_{p}}\label{eq:soundspeed}
\end{equation}
is the square of the sound speed. Further, using Eq.~(\ref{eq:radcont}) we
derive a differential equation for a single
unknown  in
a similar manner as in \citet{summers1982}
for the two fluid polytropic wind or for the classical Parker wind \citep[see,
e.g.][]{parker1958,Mann99}:
\begin{equation}
\frac{r}{v}\frac{dv}{dr}=\frac{2C_{s}^{2}-\frac{GM_{\sun}}{r}}{v^{2}-C_{s}^{2}}.
\label{eq:radmoment-4}
\end{equation}
This equation implies the existence of (a) critical (singular) point(s) in the
solution and determines its (their) position:
\begin{equation}
r=r_{c}=\frac{GM_{\sun}}{2C_{s}^{2}}=R_{\sun}\frac{V_{\sun esc}^2}{4C_{s}^{2}},
\label{eq:biffurc_position}
\end{equation}
where, the Mach number $M=v/C_s$ with $V_{\sun esc}^2={2GM_{\sun}}/{R_{\sun}}$
being the squared gravitational escape velocity at the solar surface. The family
of these transonic solutions is very well documented in the related literature
and, therefore, we will not repeat
their analysis here. We just address several fundamental aspects related to the
construction of an additional class of
solutions in the following section.

\section{A new class of transonic solutions}
\label{sec:interpretation}
\subsection{Mathematical conditions}\label{sec:newclass}
Recall that the theory of stellar winds developed by Eugene Parker was the
realization of the concept that radial variability of the gravitational force
and other physical quantities imposes an
analogy with industrial hydrodynamic sub- and supersonic nozzles and diffusers,
and Eq.~(\ref{eq:radmoment-4})
provides the mathematical background for this analogy. In the isothermal SW
model (i.e.\ with the polytropic index $\alpha=1$),
the solution space is constructed with the requirement that the nominator and
the denominator on the r.h.s.\ of
Eq.(\ref{eq:radmoment-4}) both vanish at a critical point. This requirement
permits finite values for the spatial velocity
derivative at the critical point. However, one has to notice that even when the
singularity is removed, the
presence of the bifurcation point implies {\lq}a physical/configurational
singularity{\rq}  manifested in the fact
that an observer to the right of the critical point in the $v-r$ plane may not
realize to which kind of flow pattern (sub-
or supersonic) this point belongs to, and for a clear conclusion it is needed
to move at least by an infinitesimal distance in the left or right direction
from this point (enabling a conclusion about the sign of the gradient in this
point). As a matter of fact, this ambiguity shows that sub- and
supersonic parts of the flow can represent formally mutually independent
patterns, and the position of the critical point for each
of them can be uniquely determined by the corresponding boundary conditions,
viz.\ on the left side from the critical point
these conditions are set at the base of the corona (boundary value problem) and
for the right side usually the physical
quantities at infinity define the position of the critical point.

The fact that, within the considered set up, the
integration constants in the corresponding governing equations are chosen in a
way that the positions of the critical
point derived from the left and right sides coincide, does not remove the
physical independence of these parts of the
flow pattern. This independence becomes evident if we recall the analogy with
the interconnected hydrodynamic sub-
and supersonic nozzles. As a matter of fact, it is well known that in the
subsonic nozzle the maximum Mach number that could be achieved
is $M=1$.
The same is valid for the supersonic nozzle with the difference
that the value $M=1$  represents a global
minimum in this case and it can not be dropped below this value. 

It is important to note that in this case, if one demands that
\begin{equation}
1-\frac{GM_{\sun}}{2C_{s1}^{2}r_{*}}=1-\frac{GM_{\sun}}{2C_{s2}^{2}r_{*}}=0,
\label{denom_cond}
\end{equation}
then in the critical point the nominator in
Eq.(\ref{eq:radmoment-4}) vanishes. This set of conditions restricts significantly the space of available solutions as it dictates that the governing equations must be solved subject to the just outlined conditions. It is easy to show that in such case all the solutions are smooth and for $\alpha > 3/2$ only decelerating from supersonic to subsonic states exist. The threshold value 3/2 is known as the Parker polytrope, see e.g. \citet{Tsinganos-1996}. 

However,
in general terms, the discontinuous solution is not forbidden, i.e.\ we can have a wider class of
solutions for the SW flow patterns. In general, the restriction of smooth
transonic outflow can be dropped and the nominator and denominator of the
r.h.s.\ of Eq.~(\ref{eq:radmoment-4}) can vanish at different points. In this
case the position of the singularity is determined entirely by the condition $M=1$,
which implies infinite radial gradients in velocity, density, and temperature.
In other words, subject of our paper is to explore the extention of the solution space
when the mentioned restrictions are absent, implying that there is some physical process
in the singular jump region enabling
and 
\begin{equation}
1-\frac{GM_{\sun}}{2C_{s1}^{2}r_{*}}\neq 1-\frac{GM_{\sun}}{2C_{s2}^{2}r_{*}}\neq 0,
\label{denom_cond2}
\end{equation}
at the critical point.

This way a wider class of solutions can be constructed by setting
boundary conditions accordingly, provided that
the second law of thermodynamics is respected and the entropy increases across the jump.

In order to continue the description of our formalism, it is convenient to
introduce an auxiliary function $r_{c}(r)$:
\begin{equation}
r_{c}(r)=\frac{GM_{\sun}}{2C_{s}^{2}(r)}\label{eq:biffurc_cond}.
\end{equation}
This function defines the position of the point $r_0=r_{c}(r_0)$ where the
nominator of the r.h.s.\ of
Eq.~(\ref{eq:radmoment-4}) vanishes. In case the condition (\ref{denom_cond})
holds, we have a standard
transonic smooth solution when the points $r_{01}$ and $r_{02}$ corresponding
to the "upstream" and "downstream" states satisfy the relation
$r_{01}=r_{02}=r_{*}$. Otherwise, these points differ from each other and
a jump can be formed at the critical point $r_{*}$.

In summary, the solutions for flow patterns that could be covered by
(\ref{eq:radmoment-4})
are of two kinds: (i)~when the mentioned two points are placed at the same
radial distance to maintain the existence of a
finite flow velocity derivative everywhere within the radial domain; (ii)~When
these two points are different and singular jumps in physical
quantities appear.

Further, by denoting $\eta=r/r_{c}$ as a dimensionless distance variable and
$\xi=M^2$ and using the
obvious relation
\begin{equation}
  \frac{d\ln r_{c}}{dr}=-\frac{d\ln C_{s}^{2}}{dr}, \label{rcvscs}
\end{equation}
it is straightforward to proof that the following equation is valid:
\begin{equation}
\left(5-3\alpha\right)\frac{d\ln
C_{s}^{2}}{dr}+(\alpha-1)\left(\frac{d\ln\xi}{dr}+4\frac{d\ln\eta}{dr}\right)=0,
\label{eq:radenergy3}
\end{equation}
from which it becomes evident that in the limit
$\alpha\rightarrow1$ the formalism developed here reduces to the Parker's
thermodynamically isothermal wind, and in the particular case of considered in
this paper it reduces to the constant temperature
case \citep[see, e.g.][]{Mann99} as we omit thermal conduction and/or other
transport
processes.

In order to perform a thorough analysis of the developed analytical framework,
it is convenient to present
(\ref{eq:radmoment-4}) in terms of $\eta$ and $\xi$:
\begin{equation}
\frac{d\xi}{dr}-\frac{d\ln\xi\eta^{4}}{dr}-\left(3+\xi\right)\frac{d\ln
r_{c}}{dr}=-\frac{4r_{c}}{r^{2}},\label{eq:machgeneral}
\end{equation}
which represents a most general equation generating the space of flow pattern
solutions in combination with Eq.~(\ref{eq:radenergy3}) and the boundary
conditions for the density, flow velocity and
temperature being in accordance with the above stated jump conditions.
In case of existence of a critical point above the stellar surface, the
boundary conditions are to be
selected separately upstream and downstream w.r.t.\ the critical point.

\subsection{Flow patterns of monatomic adiabatic gas $\alpha=\gamma=5/3$}
\label{sec:casemonatomic}
It is well known from the thermodynamics that monatomic gases
with intrinsic particle motions with three
degrees of freedom have the adiabatic index equal to $5/3$ (as a fraction of the specific
heat capacities $c_p/c_v$).
As noted above, we limit the study to adiabatic processes outside the 
discontinuity  to reveal the properties of the new class of solutions in a most simple
set-up. The adiabaticity implies the absence of any means for entropy generation or its
microscopic transport above the solar surface. All sources maintain the thermodynamic
state at the so-called source surface, and
the only process is the macroscopic convective transport by the flow with
velocity $v$ under the action of the gravitational force and the thermal
pressure gradient. For $\alpha=5/3$, the first term in the
Eq.~(\ref{eq:radenergy3}) vanishes and, as $\alpha \neq 1$, the solution of the
equation is:
\begin{equation}
\xi\eta^{4}=C_{*}^{2}=\mathrm{const},\label{eq:solsound5/3}
\end{equation}
where $C_{*}$ is a dimensionless integration constant (for convenience, in the
further discussion it is
introduced in the squared form). It is straightforward to show that, in case of
$\alpha=5/3$,
\begin{equation}
C_{*}=\frac{16 Q_{\rho}Q_{Cs}^{3/2}}{R_{\odot}^{2} V_{\odot
esc}^{4}},\label{Cstar5/3}
\end{equation}
where
\begin{equation}
Q_{Cs}=\frac{C_{s}^{2}}{\rho ^{\frac{2}{3}}} = \mathrm{const},\label{Qcs}
\end{equation}
is another constant of integration, representing the specific entropy, and
Eq.~(\ref{eq:machgeneral}) takes the form
\begin{equation}
\frac{d\xi}{dr}-\left(3+\xi\right)\frac{d\ln
r_{c}}{dr}=-\frac{4r_{c}}{r^{2}},\label{eq:machgeneral5/3}
\end{equation}
which governs the stationary adiabatic (isentropic) 1D flow. Notice the
substantial difference compared to its
equivalent and well-known isothermal counterpart: the term with a derivative of
$\ln \xi$ is absent, which drastically changes
 the immanent physics of the flow process!

In the isothermal case ($\alpha=1$), the condition
(\ref{eq:solsound5/3}) does not have to be satisfied and, thus, the
corresponding equation includes the term
with the derivative of $\ln \xi$ and, as a result, admits a transonic solution
which passes through the critical point
$(\xi,\eta)=(1,1)$. This solution corresponds to the solution where both the
nominator and the denominator of the
Eq.~(\ref{eq:radmoment-4})  are equal to zero (simultaneously). The latter is
achieved by the corresponding choice of the
integration constant in the equation, which physically means setting the
specific boundary conditions of
the pressure and other quantities that maintain the desirable solution (when
the Mach number becomes $=1$ in a single
critical point coinciding with the one where the escape velocity and the sound
speed values yield a vanishing nominator in
Eq.~(\ref{eq:radmoment-4})). In the case of constant temperature \citep{Mann99}
$r_{c}$ is by definition
a constant and $\eta$ thus represents just the radial distance normalized by
the critical one.
In the case of adiabatic expansion of the gas with $\alpha=5/3$, the existence
of similar solution, with
nominator and denominator simultaneously vanishing, requires $C_{*}=1$,
which again fixes the specific values of the physical quantities via
Eq.~(\ref{Cstar5/3}). It is easy to show
that, contrary to the isothermal case, now the selected solution is valid only
for the pattern for which both
$\xi \equiv 1$ and $\eta \equiv 1$ throughout of their entire domain of
definition (arbitrary $r$).

This situation can be interpreted as follows: if the specific entropy and the
mass flow rate on the source
surface and at infinity coincide, then there is no transonic solution available
as the Mach number $M=1$ everywhere. The
solution for the sound speed and flow velocity, in this case, can be written as:
\begin{equation}
v(r)=C_{s}(r)=\left ( \frac{GM_{\sun}}{2r}
\right)^{\frac{1}{2}}.\label{eq:soundvelocity5/3simple}
\end{equation}

The difference to the classic Parker's case becomes clear if we recall that the
motion of each
fluid element along the streamlines in the isothermal case is supported by some
process of microscopic transport or
generation of entropy. In the adiabatic case, there are no such transport
processes (for
instance, thermal conduction) or sources available. The specific volume
deformation of the fluid elements along their
trajectories is done entirely at the expense of internal energy, and the radial
profiles of density,
temperature and velocity are solely determined by the macroscopic convective
transport with the velocity field $v(r)$.

In order to have a single critical point in the non-isothermal cases, the
non-adiabatic (non-equilibrium) sources of the entropy production or transport
must be at work. At microscopic scales perhaps one of the good candidates again
is thermal conduction, while macroscopic MHD turbulence \cite[e.g. see][and
references therein]{vainio03,shergelashvili12} or other wave processes in
non-equilibrium media \citep{shergelashvili07,shergelashvili06} can provide a
substantial contribution into the solar wind pattern formation. The
nonadiabatic framework of the problem is beyond the scope of the present study.
Here, we just underline one important aspect of the given comparison, namely
that for both the isothermal and the adiabatic cases, for the existence of
a critical point, the nominator and denominator of
the Eq.~(\ref{eq:radmoment-4}) have to vanish simultaneously. In order to
achieve this, one has to set specific values of the physical quantities at the
boundaries. Such selection is similar to maintaining a so-called design
pressure for
the engineering realization of a subsonic-supersonic nozzle. It is plausible to
assume that apart from these
solution patterns satisfying the specifically tuned 'design pressure'
conditions, there exists a wider class
of flow solutions admitting finite nonzero values of the nominator in
Eq.~(\ref{eq:radmoment-4}) at the critical point ($M=1$) and as a result
including the cases
of the flows which have jumps (discontinuous behaviour) of the physical quantities
at that point. All those
solutions are obtained through an arbitrary choice of the values of the
integration constant from the domain
$C_{*}\neq 1$.

Let us consider this more general class of solutions for the case of
$\alpha=5/3$. By substitution of the
Eq.~(\ref{eq:solsound5/3}) into Eq.~(\ref{eq:machgeneral}) one obtains:
\begin{equation}
\frac{d}{dr} \left[ \ln\frac{\eta\left(3+C_{*}
^2\eta^{-4}\right)-4}{r}\right]=0, \label{eq:mach5/3}
\end{equation}
which translates into the algebraic quartic equation for $\eta$
\begin{equation}
3\eta^{4}-\left(4+Dr\right)\eta^{3}+C_{*} ^2=0,\label{eq:eta5_3}
\end{equation}
where $D$ is the third integration constant, the physical nature of which will
be revealed in the following
discussion. Applying the Ferrari-Cardano theory of quartic equations, the
general solution of this equation can be
written in the form:
%
\begin{equation}
\begin{array}{cc}
\eta_{1,2}= &
-\frac{b}{2a}-S\pm\frac{1}{2}\left(-4S^{2}-\frac{P}{4a^{2}}+\frac{b^{3}}{8a^{3}S
}\right)^{\frac{1}{2}}\\
\eta_{3,4}= &
-\frac{b}{2a}+S\pm\frac{1}{2}\left(-4S^{2}-\frac{P}{4a^{2}}-\frac{b^{3}}{8a^{3}S
}\right)^{\frac{1}{2}},
\end{array}\label{gensolution}
\end{equation}
%
where
%
\begin{equation}
\begin{array}{ll}
S=\frac{1}{2} &
\left(-\frac{P}{12a^{2}}+\frac{1}{3a}\left(H+\frac{\Delta_{0}}{H}\right)\right)^
{\frac{1}{2}}\\[0.2cm]
P= & -3b^{2}, \label{gensolaux1}
\end{array}
\end{equation}
and:
\begin{equation}
\begin{array}{cc}
\Delta_{0}= & 12ae\\
H= &\displaystyle
\left(\frac{\Delta_{1}+\left(\Delta_{1}^{2}-4\Delta_{0}^{3}\right)^{\frac{1}{2}}
}{2}\right)^{\frac{1}{3}}.
\label{gensolaux2}
\end{array}
\end{equation}
Here, $a=3$, $b=-(4+Dr)$ and $e=C_{*}^{2}$, $\Delta_{1}=27b^{2}e$. It is also
convenient to introduce the additional
parameters
\begin{equation}
\begin{array}{ll}
\Delta=\left(4\Delta_{0}^{3}-\Delta_{1}^{2}\right)/27=256a^{3}e^{3}-27b^{4}e^{2}
=27C_{*}^{4} \left( 256 C_{*}^{2} -
b^4 \right)\\[0.3cm]
K=64a^{3}e-3b^{4}=3\left( 576 C_{*}^{2} - b^4 \right). \label{gensolaux3}
\end{array}
\end{equation}
Our goal is to investigate  classes of solutions that give real $\eta$ values
along the considered radial
domain $r/R_{\sun}>1$.

It is evident that $P<0$. Besides, if $\Delta>0$, then $K>0$
(these relations forbid in the considered case the solution with four distinct
real solutions for any $r$). In general, these conditions significantly
restrict the set of available real solutions. To
obtain the flow patterns, the equation \label{eq:eta5/3} has to be evaluated in
each $r$,
which is subject to the boundary conditions that determine the constants $D$
and $C_{*}$. For the further analysis
it is appropriate to distinguish the following cases: (i) when $\Delta<0$, then
two out of four solutions (\ref{gensolution}) are real and distinct; (ii) when
$\Delta=0$, then multiple roots may exist. In particular, if $P<0$, $K<0$, and
$\Delta_{0}\neq0$ there are
one double and two simple real roots, when $K>0$ there exists one double real
root, and if $\Delta_{0}=0$ and
$K\neq0$ there are one triple and one simple root, all real.

We continue the analysis with the latter case from (ii) which corresponds to
$\Delta _0=36C_{*} ^2=0$ i.e. $C_{*}=0$ and, thus, no flow according to
definition (\ref{Cstar5/3}) that implies
$Q_{\rho}=r^{2}\rho v=0$. Then, Eq.(\ref{eq:eta5_3}), as expected, has one
physically meaningless triple solution
$\eta=0$ and one nontrivial solution:
\begin{equation}
\eta=\frac{4+Dr}{3}, \label{eta_solnoflow}
\end{equation}
where the constant $D\geq -4/R_{\sun}$ can be chosen arbitrarily to give the
solution for the temperature and density
profiles in the stationary streamlines:
\begin{equation}
C_{s}^{2}= Q_{Cs} \rho
^{\frac{2}{3}}=\frac{GM_{\sun}D}{6}\left(1+\frac{4}{Dr}\right)=\frac{V_{\sun
esc}^2
DR_{\sun}}{12}\left(1+\frac{4}{Dr}\right),
\label{stop_flow5_3}
\end{equation}
provided that $D>0$ as both temperature and density must be positive everywhere.

It should be noted that if the solution contains a critical point where the
$M=1$, then
the subsonic and supersonic parts of the flow have to be considered as two
interconnected states and the boundary
conditions are set such that these two states are matched in the common
critical point $r=r_{*}$ where
both states maintain $M=1$, that enabling a 'smooth' transition on the $(\xi,
\eta)$ plane. The existence of such critical point requires that the
corresponding root $\eta=\eta_{*}$ has to be
double which corresponds again to the case (ii): $\Delta=0$ with $\Delta _0
\neq 0$, which is equivalent
to $C_{*}\neq 0$ and a finite flow velocity. Then it is easy to check that
$K>0$ when $\Delta=0$. Therefore one gets a double real root, meaning that
there exists an intersection
point of the sub- to super- and super- to subsonic flow patterns which is
equivalent to the bifurcation point in the standard
Parker solutions. The only difference is that, apart from the Mach number, all
other physical quantities are
discontinuous. The latter relation immediately
gives $b=\pm 4\sqrt{C_{*}}$, so that at the critical point $\eta=\eta_{*}$
following two alternative equations are valid:
\begin{equation}
3\eta_{*}^{4}\pm 4\sqrt{C_{*}}\eta_{*}^{3}+C_{*}^{2}=0, \label{soletacr}
\end{equation}
with the two possible double roots $\eta_{*}=\mp \sqrt{C_{*}}$, respectively.
It is obvious that both roots satisfy the relation (\ref{eq:solsound5/3}) as
well.

Further, through the comparison of Eqs.~(\ref{soletacr}) and (\ref{eq:eta5_3})
one can
conclude that, accordingly, the algebraic equations having a family of
solutions that correspond to
the flows with patterns of $(\xi, \eta)$ curves passing through the critical
point $\eta=\eta_{*}$ read as:
\begin{equation}
3\eta ^{4}-4\left(1\mp \left(\sqrt{C_{*}}\pm 1\right)\frac{r}{r_{*}}
\right)\eta ^{3}+C_{*}^{2}=0,
\label{soletacrf}
\end{equation}
with
\begin{equation}
 r_{*}=\mp\frac{4\left(\sqrt{C_{*}}\pm 1\right)}{D}. \label{rstarnegativ}
\end{equation}
In the case of $\eta_{*}=- \sqrt{C_{*}}$ ($b=4\sqrt{C_{*}}$), the solution with
the critical point
$r=r_{*}$ is mathematically consistent. However, it gives negative $\eta
_{*}$-s, i.e.\ it is requiring nonphysical imaginary
sound speeds. Therefore, in this case, the part of the solution space given by
the Eq.~(\ref{soletacrf}) does not have critical
points where $M=1$ in the domain $r>0$ for any real value of the sound speed.
Consequently, this version of the equation does not generate transonic
solutions and, thus, will not be considered further. Instead, we focus on the
alternative case of Eq.~(\ref{soletacrf}) corresponding to case
$\eta_{*}=\sqrt{C_{*}}$ ($b=-4\sqrt{C_{*}}$) and will explore the core physical
content of the solutions it generates.

\subsection{Sample solutions}
\begin{figure*}
\centering
{\includegraphics[scale=0.52]{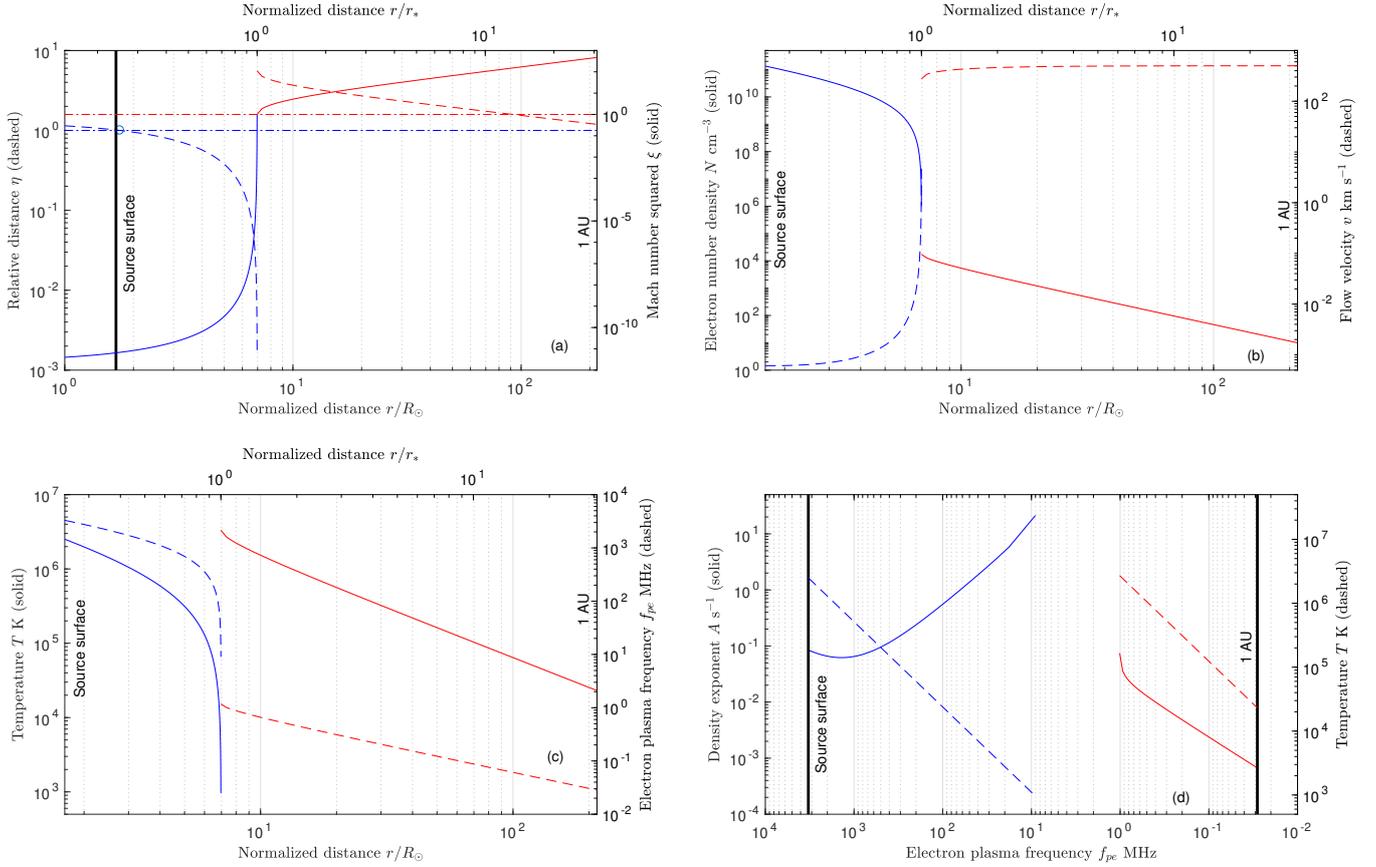}}
\caption{The slow solar wind pattern with parameters from the reference model
(SWRM) for sub (blue curves) and
supersonic (red curves) parts. Bottom $x/R_{\sun}$ and top $x/r_{*}$ (jump
region) x-axes in panels (a), (b) and (c).
The local electron plasma frequency $f_{pe}$ on x-axis in panel (d). Panel (a):
$\eta$ (dashed curve, left $y$-axis)
with horizontal blue dashed-dotted line indicating $\eta=1$ level and cycles
marking the place where the nominator of
the Eq.~(\ref{eq:radmoment-4}) vanishes. Mach number squared $\xi=M^2$ (solid
line, right $y$-axis). The red
dashed-dotted horizontal line shows the level of $M=1$. Thick vertical line -
position of the source surface $r=r^{SS}$.
panel (b) - Number density $N$ (solid line, left $y$-axis) and flow velocity
$v$ (dashed line, right $y$-axis); panel
(c) - Temperature $T$ (solid line, left $y$-axis) and electron plasma frequency
$f_{pe}$ (dashed line, right $y$-axis);
panel (d) - The number density exponent $A$ and temperature $T$ (dashed line,
right $y$-axis) against the electron
plasma frequency.}
 \label{fig_base_case1}
 \end{figure*}
\begin{figure*}
{\includegraphics[scale=0.52]{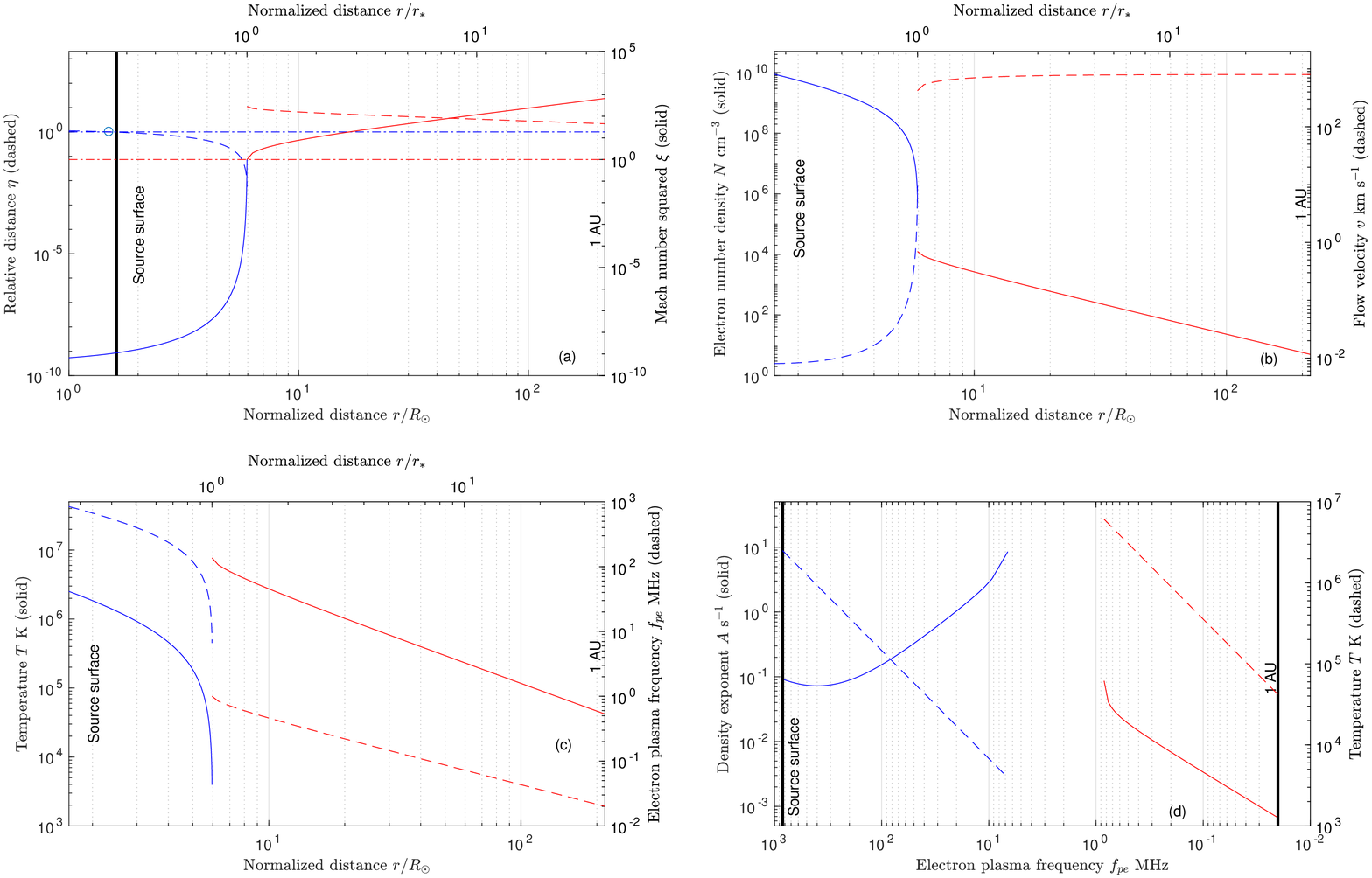}}
\caption{Same as in Figure~\ref{fig_base_case1} for fast wind typical pattern.}
 \label{fig_base_case2}
 \end{figure*}
In the present section, we perform an analysis of Eq.~(\ref{soletacrf}). Our
aim is to plot typical flow patterns corresponding to
sample sets of boundary conditions that give typical slow and fast wind
patterns. In general, the detailed investigation of the available solution
space requires a rather complex multi-variable parameter study in combination
with the observable sets of synoptic maps and in-situ measurements. However, it
should be taken into account that such a rigorous parameter study would require
the inclusion of the effects of the energy sources and other non-adiabatic
affects. While we are focusing here on a qualitative (semi-quantitative)
demonstration of the principally novel space of the solutions on the bases of
the adiabatic model, we just show two typical solutions for the slow and fast
wind patterns with input parameters (boundary conditions at source surface and
1AU) that are in accordance with both the respective solar wind categorization
schemes \citep[e.g., see][]{xu2015} and other general constraints on the
boundary conditions known from the observational data \citep{Matthaeus2006}.
These constraints form the general framework of relations between the parameter
values that enables identification (distinction) of slow and fast solar wind
patterns (in what follows we consider a mean slow wind case without subdivision
of them into strahl reversal zone and streamer belt origin patterns as is done,
for instance, by \citet{xu2015}). At any case, the categorization of the slow
and fast solar winds requires that slow wind patterns have lower
velocity and temperature and higher density values compared to the those of
fast winds. The set of input and output parameters for both illustrative
examples are given in Table~\ref{tab:ParamBC}. It is interesting to note that, brief parametric study of the solutions we build shows agreement and confirms observational evidence that the position of the sonic point is sensitive not only to the flow velocity but also to the temperature as well \citep{lotova2013}.

The solution space we are investigating covers the values of $C_{*}\neq 1$
which implies nonzero values of
parameter $D$, that itself maintains consistency and finiteness of the
Eq.~(\ref{eq:mach5/3}). In this case, a critical
point exists (like in the isothermal case) where $M\left(r_{*}\right)=1$
(denominator in Eq.~(\ref{eq:radmoment-4}) is zero). However, now $\eta
\left(r_{*}\right)=\sqrt{C_{*}}\neq 1$ (nominator in Eq.~(\ref{eq:radmoment-4})
can be non-vanishing at the same point) meaning that the stationary solutions
can sustain jumps in velocity, density and temperature radial
profiles matching at the critical point, provided that there exists some external
source of energy in that point supporting such a jump. We introduce one additional
parameter  - the
altitude $r=r^{SS}$ (measured from the center of the Sun) of the source for a
given stream and we use the superscript
$^{SS}$ for all parameter values at this altitude. Its position is determined
by the configuration of the magnetic field
structures in the corona and the position of the critical point, thus it can
change from region to region depending
which type of flow pattern is considered. As was mentioned above, in a real
situation this classification of patterns can be more complex unlike to the
simple slow-fast wind we are adopting in this paper (such categorization can
follow the one described in \citet{xu2015} or one related to 
corotating interaction regions). In both sample cases considered here, we
presume that it is situated above the solar surface $r=r^{SS}>1$.

 We conditionally define the altitude of the source surface as
the height at which the
local temperature is approximately equal to $2.0\;$MK. We also set the values of the velocity, 
temperature and density at
1~AU which satisfies two conditions (we denote them with superscript $^{AU}$):
(i) the critical points of the sub- and
supersonic parts of the flow patterns match in a single critical point thus
enabling a smooth transonic transition of
the Mach number and jumps in other quantities. (ii) the specific entropy,
enclosed in the constant $Q_{Cs}$
increases across the jump.

\begin{table}
\caption{The parameter values and boundary conditions.}
\label{tab:ParamBC}
\begin{center}
\begin{tabular}{lrr}
\hline
\textbf{Parameter} & \textbf{Slow wind} & \textbf{Fast wind}\\
 \hline
$N^{SS}$, cm$^{-3}$ & $1.35\cdot 10^{11}$ & $8.83\cdot 10^{9}$ \\
$N^{AU}$, cm$^{-3}$ & $10.0$ & $5.0$ \\
$C_s ^{SS}$, km$\cdot$s$^{-1}$ & $240.00$ & $240.00$ \\
$C_s ^{AU}$, km$\cdot$s$^{-1}$ & $23.00$ & $31.00$ \\
$T ^{SS}$, K & $2.51\cdot 10^{6}$ & $2.51\cdot 10^{6}$ \\
$T ^{AU}$, K & $2.31\cdot 10^{4}$ & $4.19\cdot 10^{4}$ \\
$V ^{SS}$, km$\cdot$s$^{-1}$ & $6.11\cdot 10^{-4}$ & $8.05\cdot 10^{-3}$ \\
$V ^{AU}$, km$\cdot$s$^{-1}$ & $500.00$ & $800$\\
$Q_{Cs} ^{SS}$, cm$^{4}$ s$^{-2}$ g$^{-2/3}$ & $2.19\cdot 10^{23}$ & $1.35\cdot
10^{24}$ \\
$Q_{Cs} ^{AU}$, cm$^{4}$ s$^{-2}$ g$^{-2/3}$ & $1.14\cdot 10^{28}$ & $3.28\cdot
10^{28}$ \\
$Q_{\rho} ^{SS}$, g s$^{-1}$ & $2.32\cdot 10^{-11}$ & $1.82\cdot 10^{-11}$ \\
$Q_{\rho} ^{AU}$, g s$^{-1}$ & $2.32\cdot 10^{-11}$ & $1.82\cdot 10^{-11}$ \\
$C_{*} ^{SS}$ & $2.61\cdot 10^{-6}$ & $3.19\cdot 10^{-5}$ \\
$C_{*} ^{AU}$ & $30.90$ & $121.07$ \\
$D ^{SS}$, cm$^{-1}$ & $-0.57$ & $-0.67$ \\
$D ^{AU}$, cm$^{-1}$ & $2.62$ & $6.72$ \\
$r^{SS}$, $R_{\sun}$ & $1.68$ & $1.61$ \\
$r_{*}$, $R_{\sun}$ & $6.97$ & $5.95$ \\
$F_e^{SS}$, erg$\cdot$ cm$^{-2}\cdot$s$^{-1}$& $-6.35\cdot 10^{3}$ & $-5.92\cdot 10^{3}$ \\
$F_e^{AU}$, erg$\cdot$ cm$^{-2}\cdot$s$^{-1}$ & $2.90\cdot 10^{4}$ & $5.95\cdot 10^{4}$ \\
$\triangle E_{tot}$, erg$\cdot$ cm$^{-3}$& $3.00\cdot 10^{-4}$ & $2.15\cdot 10^{-4}$ \\
$\left| \triangle E_{tot}/E_{1} \right|$ & $1.08$ & $1.22$ \\
 \hline
 \end{tabular}
 \end{center}
 \end{table}
The result for the slow wind pattern is shown in Figure~\ref{fig_base_case1}.
In all panels, the
blue and red colored curves correspond to the sub-and supersonic parts of the
flow patterns, respectively. On the
bottom $x$-axis of panels (a), (b) and (c) the radial distance is scaled by the
solar radius $R_{\sun}$, while on the
top horizontal axes of those panels it is normalized by
the corresponding position of the critical point $r_{*}$ (jump region). In
panel (d) physical quantities
are plotted against the local electron plasma frequency $f_{pe}=\sqrt{e^2N/\pi
m_e}$ ($e$ is the elementary
charge and $m_e$ is the electron mass). In panel (a) the range of radial
distances are
taken between solar surface and 1 AU $R_{\sun}<r<215R_{\sun}$ in order to
demonstrate the position of the source
surface
$r=r^{SS}$ (which is marked by the thick vertical black slid line), while in
the panels (b) and (c) this range
is taken
between source surface and 1 AU $r^{SS}<r<215R_{\sun}$, as we assume that the
lower boundary and
corresponding
conditions for the given wind stream is placed on the source surface and the
curves below this radial distance are
irrelevant for the current consideration. Further, in panel (a) we plot
subsonic and supersonic curves of the relative
distance $\eta$ (dashed curve, the values are shown on the left $y$-axis). The
horizontal blue dashed-dotted line
indicates the reference line at $\eta=1$ (the place where the nominator of the
Eq.~(\ref{eq:radmoment-4}) vanishes. There are
small circles marking the places of intersection of the curve of $\eta$ with
this line). The existence of such
points constitutes an important difference to the standard Parker wind 
solution, which is always a monotonically growing function of $r$,
while in our case the outflow velocity has a minimum on the subsonic and a
maximum on the supersonic side. This means that
the outflow velocity is bounded from both sides. The extremum on the subsonic
side is not relevant in our examples as it is located below the source surface 
in each case. However, the velocity maximum on the supersonic
side is very important as it means that with the set of parameters we use
starting from a certain distance, which is at most of
the order of $1\;$AU, the wind outflow should decelerate. This limits the
applicability to a finite heliocentric distance range.
The latter is, however, with up to 1 AU sufficiently large, and, thus, is 
more than needed to cover the innermost heliosphere for which we discuss the
new solutions.  
We also plot in
the same panel the  Mach number squared
$\xi=M^2$ (solid curve, the values are shown on the right $y$-axis). Again the
red dashed-dotted horizontal line shows
the level of $\xi=1$. It is clearly seen that this line intersects with the
curve of $\xi$ at jump region $r=r_{*}$
where sub- and supersonic parts of the curve meet and, consequently, there is a
smooth transition of Mach number without
jump, as expected from our set up. In the other two panels we demonstrate
corresponding radial profiles for other
physical quantities within the domain $r^{SS}<r<1\;$AU in the following
sequence: panel (b) - Number density $N$
(solid
curve, left $y$-axis) and flow velocity $v$ (dashed curve, right $y$-axis);
panel (c) - Temperature $T$ (solid curve,
left $y$-axis) and Electron plasma frequency $f_{pe}$ (dashed curve, right
$y$-axis); In panel (d) - The number
density
exponent $A$ (solid blue curve, left $y$-axis, we give detailed outline of this
parameter in section~4)
\begin{equation}
 A \sim \frac{1}{N}\frac{dN}{dr}, \label{A_defin}
\end{equation}
is reciprocal to the density scale height, and Temperature $T$ (dashed line,
right $y$-axis) are plotted vs. the electron plasma frequency (we discuss this
below). In Figure~\ref{fig_base_case2} we plot the same physical quantities
for the case of fast solar wind pattern with format and line styling identic to
that of Figure~\ref{fig_base_case1}.

Note that, while the velocity profiles shown in panel (b) of each figure 
look similar to classical Parker solutions \citep[e.g.,][]{parker1958}, there is 
a principal difference: at the discontinuity at the sonic point, there is an 
infinite gradient, i.e.\ a jump, which is not present in Parker's solution 
characterized by a smooth profile through the critical point.  

It is important to note two aspects: (i) the shown examples of both slow and
fast wind flow patterns
demonstrate that the position of the critical points and entire shape of the
patterns are extremely sensitive to the
temperature (Mach number) values and its profiles on both sides of the flow.
(ii) The curves plotted in the panel (d) of both
figures are closely related with the contemporary radio observations. 
The outline of the links with these observations will be the matter
of our forthcoming work, to be published elsewhere.

\subsection{Connection to observations}
The most insightful observational data are expected form the recently launched 
Parker Solar Probe \citep{ParkerSP2016} and Solar Orbiter \citep{SolarOrbiter2013} missions,
which will probe the solar wind source region and may discover the type of flow patterns we
model here analytically. With the advance of these missions and the increasing amount of
in-situ data one could address the specific task to check for quasistationary flow patterns
similar to those derived in the present study. 

In the meantime, until a sufficient amount of these data will be available, we can
look for possible indirect manifestations of the discontinuities in flow velocity and density
in other observations. For instance, solar radio emissions allow one to infer information on
the density, temperature and related profiles in the solar environment. The source of information
consists of the characteristic spectral properties of the different types of radio bursts in
dynamic radio spectra. Such spectra cover ranges of radio frequencies starting from
several hundreds of MHz (high density regions close to the Sun) down to tens of
kHz (interplanetary space, close to the Earth).

We demonstrate in Figures~(\ref{fig_base_case1})-(\ref{fig_base_case2}) (panels (c) 
and (d)) the ranges of the local electron plasma frequency $f_{pe}$ which, in general, constitutes
a key determining factor for the frequency spectra of the radio waves emitted by the solar wind
as well as coronal plasma and transmitted to an observer located at 1~AU. Evidently, the ranges 
of $f_{pe}$ reproduced by our solar wind model are in good agreement with observed spectra, \citep[e.g. see][]{chernov}.

The drift of the intensity peaks related to type III radio bursts enables one to determine
the mean inverted density scale-height $A$ that is also plotted as solid lines in panels (d) of
the above two figures. In addition, it is also of interest to note that $T-f_{pe}$ diagrams
(dashed lines in panels (d)) are also useful for an analysis of the model by matching it with the
temperature and density profiles inferred from the radio observations. This is one of the
directions where the results of our solar wind solutions can aim at. A similar model-dependent
analysis for the case of the isothermal solar wind has been reported earlier by
\citet{Mann99}, where the authors considered the particular case with a uniform temperature.

An important issue is how the theoretically predicted jump regions could be manifested in
the dynamic radio spectra. In general, there are few observational evidence of {\lq}a-typical{\rq} properties of radio bursts that can appear at different values of the
emitted frequency. Here we provide an example out of them.
It is intuitively evident from the obtained density profiles in the 
Figures~(\ref{fig_base_case1})-(\ref{fig_base_case2}), that the jump regions can be sources of an
rather unusual behaviour of the frequency drift rates in Type III radio bursts (in some cases, an
infinite or even changing sign drift may be observable). Observations of solar radio emission in
the decameter range revealed that occasionally there are bursts of type III that differ from
ordinary bursts of that type, particularly in the drift velocity. Typical decameter type III 
bursts have negative drift velocities, the magnitude of which varies within 2-4 MHz
s$^{-1}$ \citep{Mel2005}, but type III bursts with high frequency drift rates up to 42 MHz
s$^{-1}$ were recorded \citep{Melnikkon08} as well.

\begin{figure*}
\vspace*{1mm}
\begin{center}
\includegraphics[scale=0.35]{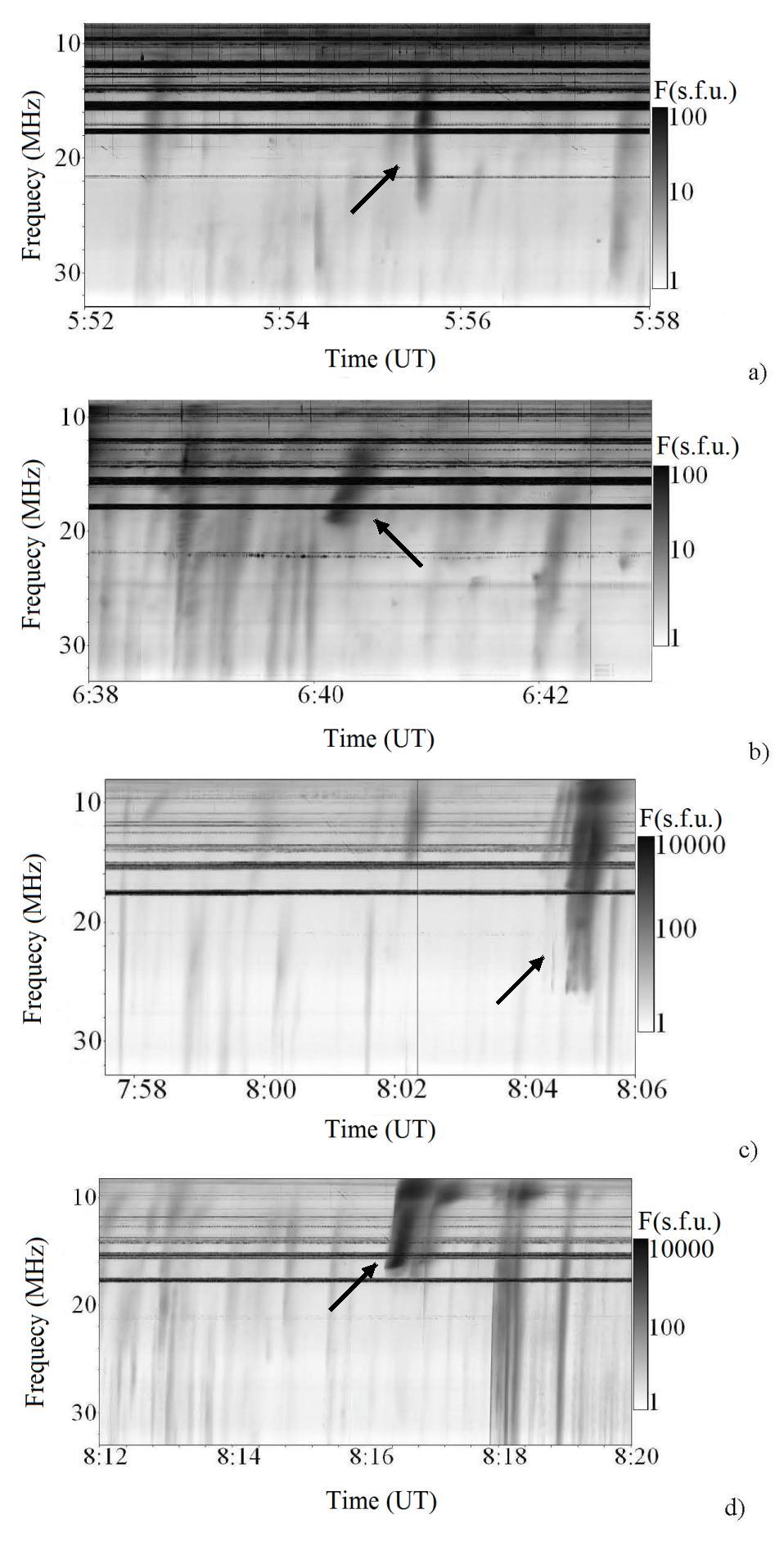}
\end{center}
\caption{Dynamic radio spectra of the caterpillar bursts observed by URAN-2 radio telescope \citep{Brazhenko2015,melnikbra2014}. The sample bursts are marked by arrows.
}
\label{unusual}
\end{figure*}

As an example, the unusual type III bursts
having a sharp, i.e.\ abrupt frequency drift breaking \citep{Brazhenko2015,melnikbra2014}
are addressed here (Figure~\ref{unusual}). In some cases, this break occurs from the high frequency side, and in some
cases from the low frequency side. Such bursts can be generated from fast electrons moving
through the coronal plasma, either before the discontinuity in the density (breaking/disappearance
from the low frequency side) or behind it, when the sudden burst occurs on the high-frequency
side.

Naturally, the development of a novel framework for the interpretation of the different
types of radio bursts on the basis of the novel analytic solutions for solar wind patterns 
obtained in this paper needs a more rigorous separate formulation and this will be carried out
in more detail elsewhere. We presented here just a single illustrative example.

\subsection{Physical conditions for the new class of solutions}
\label{heuristics}
Appearance of the derived non-smooth solutions requires a certain framework for
the constraints making them realizable in real physical situations. One 
important issue is the second law of thermodynamics that demands that the entropy
must increase along the streamlines or, at least, stays constant in isentropic flow.
A second one is the distinction of these non-smooth solutions from classic hydrodynamic
shock solutions. 

\subsubsection{Classical shocks}
In the classic case we have a supersonic flow upstream of a shock that undergoes a
strong deceleration to the subsonic state downstream. The jump occurs in a thin shock
region where it is assumed that the particle or wave-particle collision-related
dissipation (microscopic transport) processes prevail and act during the mean collision
time on mean-free-path scales. These effects are implicitly presumed in the physical
description although they are not explicitly included in the governing hydrodynamic
equations. The physics is that the kinetic energy of upstream supersonic flow is
strongly dissipated in the shock region resulting in increase of the entropy across the
shock. This process satisfies the requirement of the second law and imposes the
irreversible nature of the flow. As is known from the classic theory of shocks,
the jump of physical quantities is sudden because the fluid elements being far
upstream do not feel the presence ahead of the region with enhanced collisions.
This is due to the supersonic velocity of the ambient flow preventing acoustic waves
propagation upstream, as transmitters of the information about the pressure and
density changes in the system. Therefore, the state of the flow suddenly adapts to the
new physical conditions ahead manifested in form of the shock jump. In the SW context such
classic shock solutions have been derived for the flows with multiple critical points \citep{leer1990,holzer1977}.

\subsubsection{The new discontinuities}
In the new model the configuration of the flow and physics governing its
radial sturcture is completely different from classical shocks: the upstream flow
is subsonic and there is no kinetic energy embedded in the flow that could be
dissipated. Instead the flow is accelerated to the supersonic state formed in the
downstream region. There are two alternatives for explanation.

The first one is to consider the dynamics opposite to the one observed in the classic
case. If we allow the entropy to decrease along the streamlines, then we
obtain a picture which is forbidden by the second law, unless we take into account a more
global thermodynamic process enabling local entropy dissipation or transport to
neighboring regions. Mathematically such thermodynamical nonequilibrium process should
be manifest in the form of additional source terms in the original system of equations
(\ref{eq:contin})-(\ref{eq:energy}). However, we do not cover this possibility
with the current study.

The second possibility is to assume that the entropy is indeed increasing across the jump
at the critical point. Since such process can in our case not be sustained at the expense of
ambient flow bulk kinetic energy, some additional physical process, acting only within the
jump region of the collisionless plasma flow, is needed that could be responsible for such
a growth of the entropy. 

Among other candidate scenarios, one can draw the following possible heuristics of the
process. Let us imagine that, on the upstream side as the flow is accelerated radially the
density drops because of radial expansion and mass continuity. This means
that the plasma is adiabatically expanding and cooling, so that in a given fluid
element the temperature and, thus, the sound speed $C_s(r)$ become lower upon approaching the
critical point where $M=1$. At the same time the ambient flow velocity approaches the sonic
value, i.e.\ $v(r) = C_s(r) - \varepsilon(r)$, with $\varepsilon(r)$ denoting velocities much lower
than the sound speed. To deliver the information about the changing physical conditions to
regions far upstream backward propagating sound waves are needed. However, their effective
phase velocity in the sunward direction $\varepsilon(r)$ is very low as the flow velocity is almost sonic.

Therefore, similarly to the classic shock case, the regions far upstream are not
aware of the presence of changed physical conditions ahead. The effect is
similar but the physical reasoning behind the process is different.
Regardless of the nature of this reasoning the results is that the system is
forced to adapt to suddenly appearing physical conditions at the critical point,
providing the possibility of non-smooth jumps as described above.

\subsubsection{An estimate of required energy}
It is important to estimate the contribution of the external source of energy that can supply the jump of the flow total energy and its flux densities. For this it should be noted that from Eqs.(\ref{eq:radcont})-(\ref{eq:radenergy}) it can be readilly derived the conservation law for the total energy flux density $F_{e}$:
\begin{equation}
\frac{d}{dr}\left[\rho vr^{2}\left(h+\frac{v^{2}}{2}\right)\right]=0 \label{eq:consenergyflux}
\end{equation}
where
\[
h=-\frac{GM_{\sun}}{r}+\frac{C_{s}^{2}}{
\alpha-1}.
\]
The constant values of the total energy flux density on upstream and downstream sides differ from each other (as is shown in Table \ref{tab:ParamBC} for fast and slow wind). Therefore, The external source of energy acting in the vicinity of the critical point must produce an extra energy flux to sustain the necessary jump. The key aspect is that the total energy flux should be balanced in the critical point $r=r_{*}$, so that:
\begin{equation}
v_{1}E_{1}+C_{F_e}=
v_{2}E_{2},
\label{eq:RHnergyflux-2}
\end{equation}
where
\begin{equation}
 E_{1}=\rho_{1}\left[\frac{C_{s1}^{2}\left (\alpha+1\right )}{2\left(\alpha-1\right)}-\frac{GM_{\sun}}{r_{*}}\right], \label{energydens1}
\end{equation}
and
\begin{equation}
 E_{2}=\rho_{2}\left[\frac{C_{s2}^{2}\left (\alpha+1\right )}{2\left(\alpha-1\right)}-\frac{GM_{\sun}}{r_{*}}\right], \label{energydens2}
\end{equation}
are the total energy densities in the critical point on the upstream and downstream side, respectively, and the constant $C_{F_e}$ contains the information on the external source of energy (see further description) that supports a transition from an upstream subsonic to a downstream supersonic state. When such source of energy is absent, then the non-smooth, discontinuous flow patterns derived within our framework is not realizable and in that case solely the classic smooth Parker type transonic patterns are possible. The definitions (\ref{energydens1})-(\ref{energydens2}) can be employed to estimate the amount of total energy density that should be generated at the critical point by the external source relative to upstream total energy density (at the same point), to provide a jump-like transition from sub- to supersonic state of the flow:
\begin{equation}
\left| \frac{\triangle E_{Tot}}{E_{1}}\right|=
\left|\frac{E_{2}}{E_{1}}-1\right|,
\label{eq:RHnergy-2}
\end{equation}
The values of this relative total energy density difference are given in Table 1, accordingly. It is interesting to note that $\triangle E_{Tot}$ for both slow and fast solar wind patterns is by several orders of magnitude smaller than the ammount of total energy released by any standard flare event \citep[e.g.,][]{Fleishman2015,ellison1963}, even taking into account the radial expansion in the upward direction. This fact makes the scenarios described here easily sustainable by any reconnection type explosive event or their combinations in the solar corona.

In our case we 
observe that the temperature and density
gradients become very large at the sonic point, therefore this region can effectively
play the role of a wave reflector and, thus, serve as a source of backward propagating
acoustic waves \citep[e.g., see][]{Brekhovskikh_1,Brekhovskikh_2}. Consideration of solely longitudinal compressional 
acoustic waves is justified by two
reasons: (i) we consider here a purely radial hydrodynamic model omitting all magnetic effects; (ii) if we would
take into account the presence of the incompressible Alfv\'en waves
they propagate much faster than with sound speed, i.e.\ $V_A\gg C_s$, in
the low beta plasma and, thus, propagate out of the jump region quickly and can not
contribute efficiently to the local rapid heating and entropy generation in the vicinity of the sonic point.

For case (i), i.e.\ the reflection of acoustic waves, we can, e.g., follow the theory by \citet{Lekner-1990a} and 
\citet{Lekner-1990b} in combination with the method used by \citet{Mishonov-etal-2007}. According to these authors,
the heating power is given by a frequency integral of the damping rate $\tau_d^{-1}$ (expressed via the dissipation 
time scale) and the spectral density $\Phi(f)$:
\begin{equation}
Q_{diss}=\frac{\rho}{2}\int_{f_{min}}^{f_{max}}\frac{\Phi(f)}{\tau_{d}}df = \triangle E_{tot}
\end{equation}
where $f_{min}$ and $f_{max}$ represent the boundary frequencies of the wave spectrum.

At this point it is important to clarify from where the mentioned longitudinal (ion-) acoustic wave spectrum
can arise and whether such spectrum can be sustained by the system we consider. Evidently, the ambient flow can
take part in the formation of the spectrum of waves, but w.r.t.\ to the above modeling, we are interested in the 
situation that it is not the primary source of the wave energy. The latter is treated such that an external source is
needed. The presence of compressional longitudinal waves in such a medium is easily justified, because once the Mach
number exceeds the value 0.3, then the flow in principle can not be treated as incompressible \citep[e.g.,][]{gatski2013}.
Thus, compressional disturbances may become abundant in the flow provided that corresponding sources exist. This scenario
is further supported by the fact that both density and temperature decrease adiabatically in the subsonic part of the
flow (see our sample solutions), thereby providing a gradual growth of the compressibility coefficient $1/\rho C_s ^2$
as the flow approaches the critical point.

Consequently, upon assuming that wave sources are able to excite the acoustic modes directly or indirectly via
conversion of other wave modes, the system can sustain a certain spectrum of such waves. The direct sources include, e.g.,
magnetic reconnection related eruptive events or the existence of magnetic clouds and density inhomogeneities that play the
role of obstacles for local jet-like flows providing excitation of the waves in the wake behind the obstacle.
The indirect sources comprise, amongst other processes, weekly ( almost non-) dissipative Alfv\'en waves transporting the wave energy to the
vicinity of the critical point where they are converted into compressional waves 
\citep[e.g.][]{Jiling-1999, Malik-etal-2007}. The other mechanisms of the mode conversion can be also active, e.g. by the nonuniform magnetic field and
density (resonant absoption), or shear flows (nonmodal wave conversion). The main point is here that these alternatives, 
while operating in different regions of the solar wind source region, they do so locally. Such a local nature of the flow
pattern formation is vividly justified by the most recent observations by the Parker Solar Probe mission that already revealed a complex structure of the solar wind
\citep{Bale-etal-2019, Kasper-etal-2019}.

It is known that when the electron and ion temperatures are equal $T_e\approx T_i$, then the ion-acoustic waves undergo
strong Landau damping and their phase velocity $C_{si} ^2=k_B T_e /m_i$ is close to the ion thermal velocity $k_B T_i /m_i$
where the damping rate is maximum. Hence, the relation between the propagation and damping time scales should accurately be
taken into account. The following scenarios maybe realized locally in the soure region: (i) When $T_e \neq T_i$ it is 
expected that the ion-acoustic waves have a finite lifetime and are able to propagate towards the critical point, get
reflected and damp; (ii) When the Landau damping is strong ($T_e \approx T_i$) the waves get dissipated within a wave
period. To make this process efficient for the temperature jump at the critical point, the sources must be situated in the
immediate vicinity of or even within the critical point.

\subsubsection{An estimate of the intensity of refelcted waves}
The goal is estimating the wave energy needed to sustain the sharp temperature jump in the total energy density $\triangle E_{tot}$. 
Following \citet{Mishonov-etal-2007} the damping rate can be expressed as $\tau_d^{-1} = \nu_{tot} k^2$ with the total
'viscosity' $\nu_{tot}$ and the wave number $k$, which for the reflected waves with phase speed $\varepsilon$ and 
frequency $f$ is $k = 2\pi f/\varepsilon$. The total viscosity may include three types of transport processes, i.e.,
$\nu_{tot}=\nu_{k}+\nu_{wp}+\nu_{ww}$, where the three terms denote the kinematic, wave-particle and wave-wave 'viscosities',
respectively. While $\nu_{k}$ can arise locally due to particle collisions, $\nu_{wp}$ and $\nu_{ww}$ can be due to the 
Landau damping of the waves and wave turbulence, respectively.
Upon considering a power law spectrum $\Phi(f)=K f^{-x}$ for the reflected waves, we obtain
\begin{equation}
\frac{K\rho\varepsilon^{2}}{8\pi^{2}\nu_{Tot}(x+1)}\left(f_{min} ^{-(x+1)}-f_{max} ^{-(x+1)}\right)=\triangle E_{tot}
\end{equation}
From this equation one can, for a given temperature jump $\triangle T$, estimate the constant $K$ as a measure of the reflected wave field intensity.

So, in general terms, instead of the kinetic energy of the flow, the dissipation of acoustic waves can be the
source of the entropy and heat, i.e.\ particle acceleration in the jump region. The produced heat leads to a
drastic increase of the temperature and to local non-adiabatic expansion of the plasma, resulting in the
discontinuous rarefaction jump in density. This scenario is consistent with the second law and it is what
we observe in our example solutions.

\section{Summary and conclusions}
In this paper we reported on the results of a new, rather simplified, one-dimensional
quasi-polytropic radial solar wind expansion model, with particular emphasis on the case
of an adiabatic index $\alpha =5/3$. We demonstrated that, in this case, the governing
equations reduce to a specific form generating exact analytic solutions for the
solar wind patterns. The solutions respect both mathematical and physical
consistency conditions and thus could be used for the numerical modelling of
the solar wind using also observed solar synoptic maps for better prediction of
the space weather conditions.

The analytic framework derived in this study provides a substantial novel
class of solutions, which have not been addressed in any previous model of the
solar wind. The novelty in the solution space occurs as a result of a local heating 
enabling (provided that Mach number is still continuous) jumps of the density, 
velocity, and temperature at the sonic point.
Rather than modelling such heating explicitly, it is implicit in the construction of the solution by matching appropriate sub- and supersonic branches, this way allowing for a simplified modelling. The similar framework can be used not only for the large radielly extended solar wind patterns, but also for the modeling of small-scale flows in solar corona e.g. coronal jets \cite[see,][and references therein ]{Bagashvili18}.

There are other immediate extensions of this study. For instance the
generalization for the case of arbitrary value of the polytropic index, also
including different types of sources of the heat and acceleration of the solar
wind plasma. The model can be used for the construction of more complexly
structured multidimensional analytical solar and stellar wind patterns
\citep[see, e.g.][and references therein]{Sauty-etal-1998, Sauty-etal-2002,
Tsinganos-2007}. Thus, the
theoretical model demonstrated in this manuscript could be considered as 
paving the way towards a development of an increased variety of solar wind models 
useful for space weather and/or heliospheric studies.
\section*{Acknowledgements}
The work was supported by Shota Rustaveli National Science Foundation grants
DI-2016-52, FR17\_609 and partially by European Commission FP7-PEOPLE-2010-IRSES-269299 project - SOLSPANET. The work of G.D. has been realized within the framework of Shota Rustaveli National Science Foundation PHD student's research (individual) grant DO/138/6-310/14 and grant for young scientists for scientific research internship abroad IG/47/1/16. Work of B.M.S. was supported by the Austrian Fonds zur Forderung der wissenschaftlichen Forschung (FWF) under projects P25640-N27, S11606-N16. B.M.S. and H.~Fichtner acknowledge support from the German Research Foundation (DFG) via the collaboration grant FI~706/25-1. Work of B.M.S. and H.~Foysi was partially supported by the DFG collaboration grant FO~674/12-1. The work of T.V.Z. was supported by FWF under projects P25640-N27 and P30695-N27. M.L.K. additionally acknowledges the support by FWF project I2939-N27 and the grant No. 18-12-00080 of the Russian Science Foundation, as well as the project "Study of stars with exoplanets" within the grant No.075-15-2019-1875 from the government of Russian Federation. S.P. was supported by the projects GOA/2015-014 (KU Leuven), G.0A23.16N (FWO-Vlaanderen) and C~90347 (ESA Prodex). We are thankful to the anonymous reviewer for constructive and valuable remarks on our manuscript that led to significant improvement of the content.



\bibliographystyle{mnras}
\bibliography{mybibg} 


\bsp	
\label{lastpage}
\end{document}